\documentclass[journal=jacsat,manuscript=article]{achemso}

\usepackage[version=3]{mhchem} % Formula subscripts using \ce{}
\usepackage{fontawesome}
\usepackage{times}
\usepackage{braket}
\usepackage{txfonts}
\usepackage{amsmath}
\usepackage{amssymb}
\usepackage{caption}
\usepackage{subcaption}

\providecommand{\keywords}[1]
{
  \small	
  \textbf{\textit{Keywords---}} #1
}

\author{Yi Sun}
\affiliation[UChicago]
{Department of Chemistry, Chicago Center for Theoretical Chemistry, James Franck Institute, and Institute for Biophysical Dynamics, The University of Chicago, Chicago, Illinois 60637, United States}
\email{ys327@uchicago.edu}

\title{The Calculation of the Rate Constants via a Master Equation approach -- A Comprehensive Tutorial}


\date{\today}

\begin{document}

\begin{abstract}
Intermolecular hydrogen transfer free radical reactions are common in the combustion process and in a number of organic chemistry reactions. Therefore, evaluating the pressure and temperature-dependent rate constants of them is of great importance. Here, we present a tutorial on how a Master Equation model can be constructed to evaluate the rate constants of reversible hydrogen transfer reactions that involve tunnelling, hindered rotor effects, as well as source and depletion. 
\end{abstract}

\maketitle
\raggedbottom

%%%%%%%%%%%%%%%%%%%%%%%%%%%%%%%%%%%%%%%%%%%%%%%%%%%%%%%%%%%%%%
\section{Introduction}
%%%%%%%%%%%%%%%%%%%%%%%%%%%%%%%%%%%%%%%%%%%%%%%%%%%%%%%%%%%%%%%%%%%%%%%%%%%%
\

Intermolecular hydrogen abstraction (I-HAT) reactions are essential in organic chemistry\cite{Curan}. A prime example is the Barton reaction\cite{Barton1}, a photochemical reaction that transforms an alkyl nitrite to a $\delta$-nitro alcohol. One key step in its mechanism (shown in Figure 1) involves a hydrogen atom transfer via a six-membered ring transition state\cite{IUPAC}. Such an internal hydrogen-atom abstraction process can also occur in acetals and ethers \cite{Cosme,Zlot,Guo}.

\begin{figure}[H]
\includegraphics[width=\linewidth]{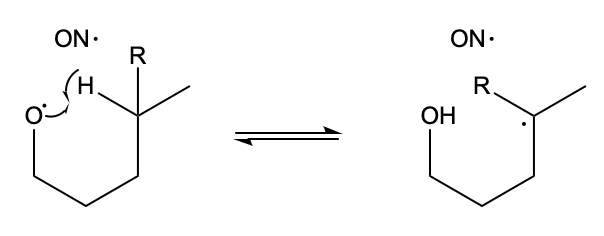}
\caption{A key step of the Barton reaction}
\centering
\end{figure}

There are also numerous examples of gas-phase I-HAT reactions, one of which lies in the oxidation process of cyclobutyl methyl ketone \cite{Prap}. Another example could be the process of the oxidation by SO2 of \emph{o}-xylene and \emph{p}-xylene, or the atmospheric photochemical oxidation of benzene \cite{Sinha}. Therefore, constructing a model to calculate the pressure and temperature rate constants is of significant interest.

Traditional models are typically based on transition state theory, in which all the species are in a Boltzmann distribution\cite{Callear}. However, when the pressure is low, the number of collisions that occur between molecules is too infrequent to exchange molecular energies sufficiently fast to compete with chemical reaction steps, leading to the well-known fall-off behavior of unimolecular rate constants\cite{Gilbert}. There are also tunnelling effects due to the fact that hydrogen is a light nucleus \cite{Anslyn}. In the kinetic isotope effect, tunnelling may lead to a 25:1 ratio in $\frac{k_H}{k_D}$ \cite{Lewis}. Additionally, the hindered rotors in the reactants may affect the rotational partition function. There have been numerous packages that support the calculation of rate constants based on these considerations, including Multiwell\cite{Barker}, MESMER \cite{David}, and MESS\cite{Jasper}, making the calculation of rate constants using Master Equation approaches routine. There have also been numerous developments and applications of RRKM theory to study combustion, which contains numerous amounts of free radical reactions\cite{Miller,Georg,Stephen}.

The goal of this paper is to provide a simple tutorial on how the basics of the Master Equation can be used to model chemical reactions of interest. To achieve this, we divide the paper into several parts, including constructing a prototypical Master Equation model from scratch, and how it could be modified to include tunnelling, hindered rotor effects, as well as source and depletion. 
%%%%%%%%%%%%%%%%%%%%%%%%%%%%%%%%%%%%%%%%%%%%%%%%%%%%%%%%%%%%%%%%%%%%%%

\section{Basic RRKM Theory and the Master Equation Model}

\subsection{Basic RRKM Theory}
%%%%%%%%%%%%%%%%%%%%%%%%%%%%%%%%%%%%%%%%%%%%%%%%%%%%%%%%%%%%%%
\

The basic result of the RRKM theory shows that the rate constant equals
\begin{equation} \label{eq:4}
k(E)=\frac{\int_{0}^{E-E_0}\rho^{+}(E^+)dE^+ }{h\rho(E)}
\end{equation}
where $E^+$ is the amount of energy out of the reaction coordinate, $E_0$ is the barrier height, and $\rho^{+}(E)$ is the density of states with the reaction coordinate removed \cite{Marcus}.

The goal is then to find an expression for the density of states. To achieve this, it can be noted that the density of states $\rho$ and the partition function $q$ are connected via
\begin{equation} \label{eq:1}
q(\beta)=\int_{0}^{\infty} \rho(E)e^{-\beta E}dE
\end{equation}
which is a Laplace transform. Therefore
\begin{equation} \label{eq:2}
q(\beta)=\mathcal{L}\left\{\rho(E)\right\}
\end{equation}
\begin{equation} \label{eq:3}
\rho(E)=\mathcal{L}^{-1}\left\{q(\beta)\right\}
\end{equation}

\subsection{The Master Equation Model}

\

As the Boltzmann distribution is not satisfied at low pressures due to the infrequent nature of molecular collisions, a detailed method is required to determine the system's energy distribution. Here, instead of using the strong collision model, which assumes that the goal of each collision is to achieve the Boltzmann distribution, we use the weak collision model to derive the matrix form of the master equation. The result is
\begin{equation} \label{eq:5}
\frac{\text{d}\textbf{p}}{\text{d}t}=(\omega(\textbf{P}-\textbf{I})-\textbf{K})\textbf{p}=\textbf{M}\textbf{p}
\end{equation}
where $\omega$ is the collision frequency, element $\textbf{P}_{ij}$ in $\textbf{P}$ is the probability to transfer from the $j_{th}$ energy grain to the $i_{th}$ energy grain in a single collision, $\textbf{I}$ is the identity matrix, $\textbf{K}$ is a diagonal matrix containing the microcanonical rate coefficients calculated from the equations (1) and (2) in Section II averaged over each grain, and $\textbf{p}$ is the vector that represents the population distribution of the reactant.

The next task is to find a way to construct the transition probability matrix $\textbf{P}$ between energy grains. Although the energy of the molecule does not have an upper limit , in reality, an upper bound is set here to make the size of the matrix finite. As long as it is sufficiently high that its Boltzmann population is negligible, the position of the cut-off does not make any sensible difference.

The column sum of this matrix must be normalised as it is a discrete probability distribution, which can be expressed as
\begin{equation} \label{eq:6}
\sum_{j=j_{min}}^NP_{ji}=1
\end{equation}

For downward transitions, an exponential model is used, which assumes that the probability is an exponentially decreasing function of the energy transferred, so that if $j \leq i$,
\begin{equation} \label{eq:7}
\textbf{P}_{ji}=A_{i} \text{exp}\left(-\alpha(E_i-E_j)\right)
\end{equation}
where $A_i$ is the normalisation constant and $\alpha$ is empirically set to $\frac{1}{200} cm^{-1}$. The expression of upward transition probabilities and normalisation constants can then be derived based on equations (6), (7), and the detailed balance expression
\begin{equation} \label{eq:8}
\textbf{P}_{ji} \textbf{p}_i=\textbf{P}_{ij} \textbf{p}_j
\end{equation}

It can be derived that 
\begin{equation} \label{eq:8b}
P_{j i}=A_j \frac{\rho_j}{\rho_i} e^{-(\alpha+\beta)\left(E_j-E_i\right)}(j>i)
\end{equation}
and 
\begin{equation} \label{eq:8c}
A_i=\frac{1-\sum_{j=i+1}^N A_j \frac{\rho_j}{\rho_i} e^{-(\alpha+\beta)\left(E_j-E_i\right)}}{\sum_{j=j_{\min }}^i e^{-\alpha\left(E_i-E_j\right)}}
\end{equation}
in which $\rho_{i}$ and $\rho_{j}$ represents the density of states in grain i and grain j, respectively if a maximum energy grain $N$ is set such that only downward energy transitions are possible.

Another unknown quantity in Master Equation (5) is the corrected collision frequency, $\Omega$. The hard sphere model suggests that
\begin{equation} \label{eq:9}
\omega=\frac{N}{V}\pi \sigma^2 \overline{v_{rel}}
\end{equation}
in which $\sigma$ is the collision radius, i.e., the sum or the radii of the colliding particles, $\overline{v_{rel}} $ is the relative speed, and $\frac{N}{V}$ is the concentration of molecules.

Repulsive interactions dominate molecular collisions, making the hard sphere model quite accurate as molecules are essentially impenetrable like hard spheres. However, it underestimates the collision frequency as it fails to account for Van der Waals interactions, and real molecules are not hard spheres. One can correct this using a modified cross-section known as the reduced collision integral \cite{Neufeld},
\begin{equation} \label{eq:10}
\Omega=\pi \sigma^2 \Omega^{(2,2)*}(T^*)
\end{equation}
where $T^*=\frac{kT}{\epsilon}$. Both $\sigma$ and $\epsilon$ are Lennard-Jones parameters.

\subsection{The Reversible Master Equation Model}

\

By coupling two MEs of isomerisable species A and B and concatenating the two population vectors into a single vector,
\begin{equation} \label{eq:11}
\frac{\text{d}\textbf{p}}{\text{d}t}=\frac{\text{d}}{\text{d}t}\begin{pmatrix}
\omega^A(\textbf{P}^A-\textbf{I}^A)-\textbf{K}^{AB} & \textbf{K}^{BA} \\
\textbf{K}^{AB} & \omega^B(\textbf{P}^B-\textbf{I}^B)-\textbf{K}^{BA} \end{pmatrix} \left(\begin{array}{c}\textbf{p}^A\\ \textbf{p}^B\end{array}\right) = \textbf{M}\textbf{p}
\end{equation}
in which $\textbf{P}^A$ and $\textbf{P}^B$ are the energy transfer matrices for the two isomers, and $\textbf{K}^{AB}$ and $\textbf{K}^{BA}$ are the diagonal matrices containing the microcanonical rate constants for the reactions from A to B and from B to A, respectively. The microcanonical rate constants can only link a grain in one isomer with the isoenergetic grain of the other isomer due to the principle of conservation of energy.

Then the relaxation rate constant can be found by diagonizing the matrix $\textbf{M}$, yielding a number of eigenvalues. The smallest eigenvalue $\lambda_0$ is zero because the column sums of the matrix M are zero, which represents the equilibrium case. The negative value of the second-smallest eigenvalue, $-\lambda_1$ is the relaxation eigenvalue, which dictates the system’s approach to equilibrium. As this eigenvalue corresponds to the chemical relaxation rate of the elementary reversible reaction scheme,
\begin{equation} \label{eq:12}
-\lambda_1 = k_{AB}+k_{BA}
\end{equation}

\subsection{Obtaining the numerical parameters}

\

In this study, all of the theories mentioned will be applied to a hydrogen transfer reaction, which is shown in Figure 2.
\begin{figure}[H]
\includegraphics[width=\linewidth]{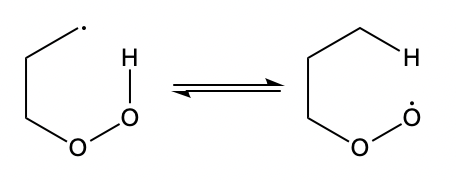}
\caption{A step in the combustion process of propane}
\centering
\end{figure}
The reaction participates in the oxidation, i.e., combustion pathway of propane, as the reactant forms through the free radical addition reaction between triplet oxygen and the 1-propyl radical. It should be noted that this reaction has been comprehensively researched in a multitude of papers\cite{DeSain1,DeSain2,Huang,GoldSmith,Slagle,DeSain3}, and the goal of this tutorial is to provide a simple yet effective approach to building a Master Equation model to study it.

The barrier height $E_0$ is essential in using RRKM to model chemical reactions, as it will be used in the RRKM expression (1). For the simplicity of this tutorial, all the quantum calculations are done using Gaussian09 at the DFT/B3LYP/6-31G*(d,p) level of theory \cite{Gaussian}. It is clearly worth noticing that more modern theories, including DFT-D and Coupled Cluster theory, can give more accurate results\cite{Grimme1,Grimme2,Grimme3,Rag}. The transition state is located using the QST2 algorithm, and the barrier height is determined by finding the energy difference between the reactant (product) and the transition state. All the structures are checked using frequency analysis, since the transition state should only contain one imaginative frequency. The value of the barrier height and the value of the imaginative frequency are used in the RRKM subroutine and the tunnelling subroutine.

Another parameter that should be calculated beforehand is the collision frequency $\omega$. As the bath gas in this simulation is set to be Argon, the collision radius is approximated by $\sigma \approx \frac{1}{2}(\sigma_{Ar}+\sigma_{radical})$, and $\epsilon$ is approximated by $\epsilon = \sqrt{\epsilon_{Ar} \epsilon_{radical}}$. The temperature is set to 600K.

When it comes to modelling the collision integral, as the analytical form of the reduced collision integral is extremely complex, a simple form obtained from curve fitting by Neufeld \cite{Neufeld} is used:
\begin{equation} \label{eq:23}
\Omega^{(2,2)*}(T^*) \approx \frac{A}{T^{*B}}+ C e^{-DT^*}+E e^{-FT^*}
\end{equation}

\subsection{Relevant Experimental Results and Explanation}

\

The most significant insight is a comparison between using irreversible ME and reversible ME in calculating the rate constants. Note that in the reversible case, the forward and reverse rate constants can be determined by solving the coupled equations.
\begin{equation} \label{eq:24}
\begin{cases}\frac{k_{AB}}{k_{BA}}=K_{eq} \\ k_{AB} + k_{BA} = -\lambda_1\end{cases}
\end{equation}
The rate constants are plotted in Figure 3.
\begin{figure}[htbp]
\includegraphics[width=\linewidth]{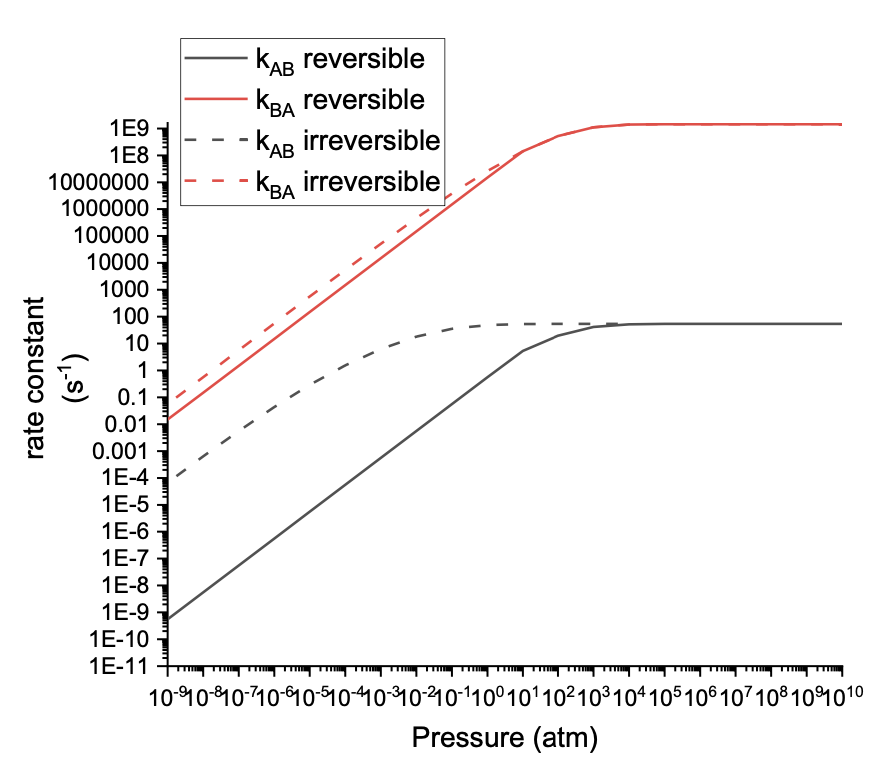}
\caption{The difference in the rate constants at 600K}
\centering
\end{figure}

According to this graph, although the deviation for $k_{BA}$ is not huge, $k_{AB}$ is. The reason behind this phenomenon is that the reaction barrier is higher (around 10300 wavenumbers) for the forward reaction but rather low (around 3000 wavenumbers) for the reverse reaction. Therefore, assuming the forward reaction is irreversible is unreasonable and will lead to highly inaccurate results in the fall-off region.

There is clearly a very large discrepancy between the rate constants deduced from the relaxation rate and the equilibrium constant and those calculated assuming that the reaction is irreversible at low pressures. This arises because in an irreversible reaction, the fall-off simply arises from the reactive depletion of the energised levels, and at low pressures, the collisional process is not sufficient to sustain Boltzmann populations. In a reversible reaction, such a depletion does take place, but it is counteracted by the reverse reaction, which leads to a repopulation of the energised states. The reaction is not complete until the energized molecules are deactivated in the isomers' Boltzmann wells. The reaction is highly endoergic, which means that the reverse reaction is much faster than the forward reaction. In the reverse reaction, energised molecules react rapidly to make the peroxyl radical, and the replenishment of these levels is slow because the microcanonical forward rate constant is small, so there is only a relatively small effect on the rate constant; nonetheless, it is significant (a factor of about 2). For the forward reaction, the depleted levels are immediately replenished by the fast reverse reaction, leading to a huge reduction in the forward rate constant, as energised molecules are much more likely to return to the reactant than to be deactivated as the product.

\section{The inclusion of tunnelling}

\

The reason why the tunnelling effect may be important is that the reaction that is studied might involve light atoms. If the hydrogen atom is transferred in the reaction, the quantum tunnelling effect will be significant and should not be ignored. One of the assumptions of RRKM theory, which states that the motion along the reaction coordinate is classical, fails if the tunnelling correction is considered. To deal with this problem, the simplest way to include quantum effects is to allow the motion along the reaction coordinate to become quantum without relaxing any of the other assumptions of TST, notably still assuming the existence of a unique reaction coordinate, which is separable from all the other motions, but including a tunnelling correction.

Eckart potential is usually used to tackle quantum mechanical tunelling\cite{Eckart}. The asymmetric form of the potential was used because of the energy difference between the two isomers. An advantage of the Eckart potential is that the solution to the time independent SE to this potential is known analytically, and the tunnelling probability is relatively easy to calculate explicitly \cite{Eckart}.

The Eckart potential \cite{Eckart} has the form
\begin{equation} \label{eq:13}
V = -\frac{Ay}{1-y}-\frac{By}{(1-y)^2}
\end{equation}
in which
\begin{equation} \label{eq:14}
y=-\text{exp}\left(\frac{2\pi x}{L}\right)
\end{equation}

The value of $x$ represents the position of the system along the reaction coordinate, and $A$, $B$ and $L$ are all parameters that can be found by fitting the potential energy curve and the single imaginary frequency of the transition state. In short, $A$ represents the energy difference between reactant and product, $L$ measures the barrier width, and $B$ is a measure of the barrier height.

The solution of the Eckart potential, hence the transmission probability,
\begin{equation} \label{eq:15}
p_{tra}(E-E^+)=1-\frac{\cosh{2\pi (\alpha-\beta)}+\cosh{2\pi \delta}}{\cosh{2\pi (\alpha+\beta)}+\cosh{2\pi \delta}}
\end{equation}
in which $\alpha$, $\beta$ and $\delta$ can all be expressed in terms of $A$, $B$, energy, barrier height, and the negative frequency of the transition state \cite{Eckart}.

The RRKM rate expression should also be adjusted; in a quantum description, there is a tunnelling probability for energies below the reaction threshold and a non-zero reflection probability for energies above the reaction threshold. The minimum value of $E^+$ is still zero (all energy in the reaction coordinate), but the maximum value is E (zero energy in the reaction coordinate).

This gives the quantum-corrected TST rate constant as
\begin{equation} \label{eq:16}
k(E)=\frac{\int_{0}^{E}\rho^{+}(E^+)p_{tra}(E-E^+)dE^+}{h\rho(E)}
\end{equation}

\subsection{Relevant Experimental Results and Explanation}

\

The changes in the RRKM rate constants are shown in Figure 4.
\begin{figure}[H]
\includegraphics[width=\linewidth]{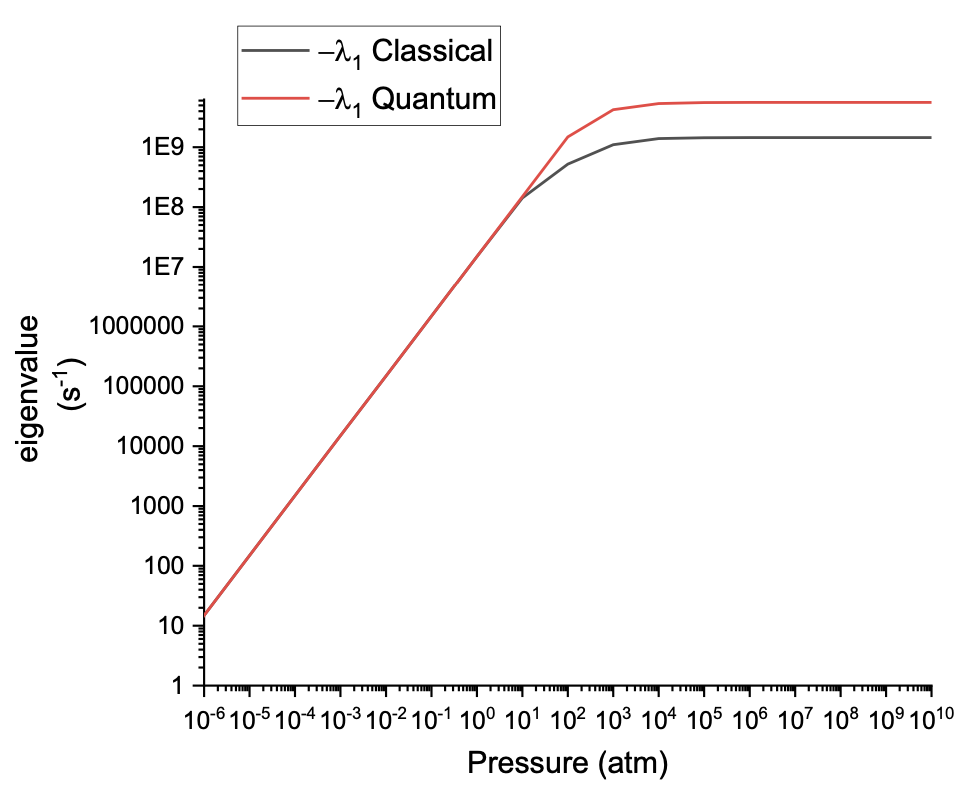}
\caption{The relaxation rate constant with tunnelling at 600K}
\centering
\end{figure}

According to this graph, the quantum tunnelling effect will also increase the forward and reverse rate constants deduced for the reversible reaction. The effect of tunnelling is more pronounced at high pressures while being negligible under low pressures, where the rate-determining step is collisional deactivation of the energised molecules, which have reached a microcanonical equilibrium.

\section{The inclusion of hindered rotors}

\

The reason why hindered rotor corrections may be important is that the molecule is an asymmetric top. According to the laws of classical mechanics, a non-linear molecule will possess three external rotational degrees of freedom. These are free rotations with no potential energy associated with them. However, many molecules also have internal rotational degrees of freedom. These have an associated potential energy, and the internal rotation is hindered, so that it may be a torsional vibration at low energies and an internal rotation at higher energies.

Despite the significance of internal degrees of freedom, they are often characterised as low-frequency vibrations when calculating the molecule’s normal coordinates. This approximation is fine at low temperatures but will break down at high temperatures. The breakdown can be seen to be significant because the high temperature limit of a vibrational partition function (assumed harmonic) is proportional to $T$, and for a rotational degree of freedom, the high temperature form is proportional to $T^{\frac{1}{2}}$. For a vibration, the densities of states increase more rapidly with energy than for a rotation. To tackle this problem, it is required to single out these low-frequency vibrations and characterise them as internal rotations. To reflect this in a calculation, the way of evaluating the partition function, therefore the density of states of the molecule, should be adjusted. While it can be assumed that the internal rotations and the external rotations can be separated, this is, in fact, an approximation since the internal rotations may affect the external moment of inertia. However, for the sake of simplicity in this tutorial, we assume the separability of these rotations.

For a hindered internal rotor, the classical partition function is
\begin{equation} \label{eq:17}
q_{int}=\frac{1}{\sigma}\sqrt{\frac{I_{int}}{2\pi \beta \hbar^2}} \int_{0}^{2\pi}e^{-\beta V(\theta)}d\theta
\end{equation}
where $\sigma$ is the symmetry number, $I_{int}$ is the effective moment of inertia, and $V(\theta)$ is the function of the barrier height from the dihedral angle.

An expression derived by Frenkel \cite{Frenkel}, et.al is
\begin{equation} \label{eq:18}
I_{int}=I_1 - I_1^{2}\left(\frac{\alpha^2}{I_A}+\frac{\beta^2}{I_B}+\frac{\gamma^2}{I_C}\right)
\end{equation}
in which $I_1$ is the moment of inertia of a group connected to the bond, $I_A$, $I_B$ and $I_C$ are the moments of inertia of the principal axis, $\alpha$, $\beta$ and $\gamma$ are the cosines of the angles between the axis of the internal rotation and each principal axis A, B, and C of the corresponding external rotation.

Suppose that the full rotational range 0 to $2\pi$ is divided into $k$ parts labelled $n$ from 1 to $k$. Each part connects one potential turning point to the next. In each part, the potential has a minimum at energy $V_{0n}$ and a maximum at energy $V_{1n}$ . In each part, we approximate the potential using the functional form
\begin{equation} \label{eq:19}
\sum_{n=1}^k\left[\frac{1}{2}(V_{1n}+V_{0n})+\frac{1}{2}(V_{1n}-V_{0n})\cos{a(\theta-\theta_{1n}})\right]
\end{equation}
where
\begin{equation} \label{eq:20}
a=\frac{\pi}{\theta_{2n}-\theta_{1n}}
\end{equation}
This can be done in practice by performing a scan calculation on three (and four) bonds with respect to the reactant and the product. 

It can be shown that the integral in equation (\ref{eq:17}) equals to
\begin{equation} \label{eq:21}
\sum_{n=1}^k\frac{e^{-\frac{\beta}{2}(V_{1n}-V_{0n})}}{a}\times \text{I}_0\left[\frac{\beta}{2}(V_{1n}-V_{0n})\right]
\end{equation}
where $\text{I}_0$ represents the modified Bessel function of order 0.

The inverse Laplace transform of equation (\ref{eq:21}) can be found using Roberts and Kaufman Tables of Laplace Transforms eq 1.15 (p 170) and eq 12.3.1 (p 257) \cite{Roberts}, which equals to
\begin{equation} \label{eq:22}
\frac{1}{\pi a}\sum_{n=1}^k\frac{1}{\sqrt{(E-V_{0n})(V_{1n}-E)}}
\end{equation}

Hence, it is easy to calculate the internal rotational partition function and the configurational contribution to the density of states. To calculate the overall density of states of a molecule, one must convolute these densities of states with the remaining vibrations according to the convolution theorem. By also considering the tunnelling effect, the pressure-dependent rate constants can be found by finding the negative value of the second smallest eigenvalue of the matrix $\textbf{M}$.

\subsection{Relevant Experimental Results and Explanation}

\

The expected difference in the relaxation rate constants is plotted in Figure 5.
\begin{figure}[H]
\includegraphics[width=\linewidth]{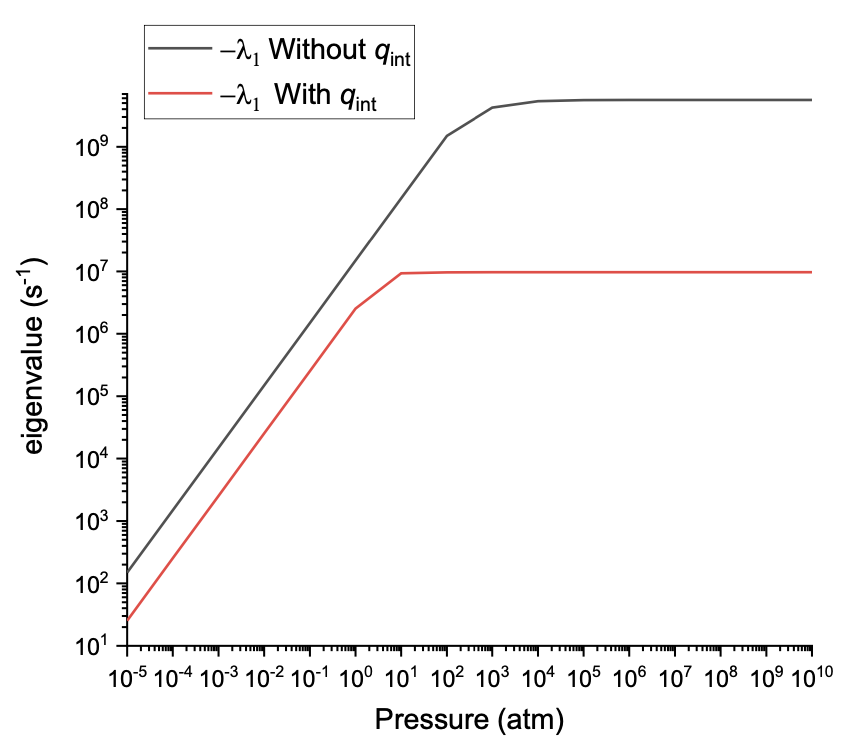}
\caption{The relaxation rate constant with internal rotors at 600K}
\centering
\end{figure}

According to this graph, it is clear that the existence of internal rotors will significantly reduce the magnitude of the rate constants deduced for both forward and backward reactions. This can be explained according to the principles of statistical mechanics, as the constraint that the transition state can only exist when there is essentially no energy in the lowest-frequency motions is a major correction.

\section{Source and depletion}

\

This section is about how source and depletion in a reversible reaction can be included in RRKM-ME modelling by expanding the Master Equation Matrix. More details are provided as including source and depletion is far less trivial when compared to other modifications. 

For the reaction shown in Figure 2, it can be extended to include a source and potential depletion routes, as shown in Figure 6 below:
\begin{figure}[H]
\includegraphics[width=\linewidth]{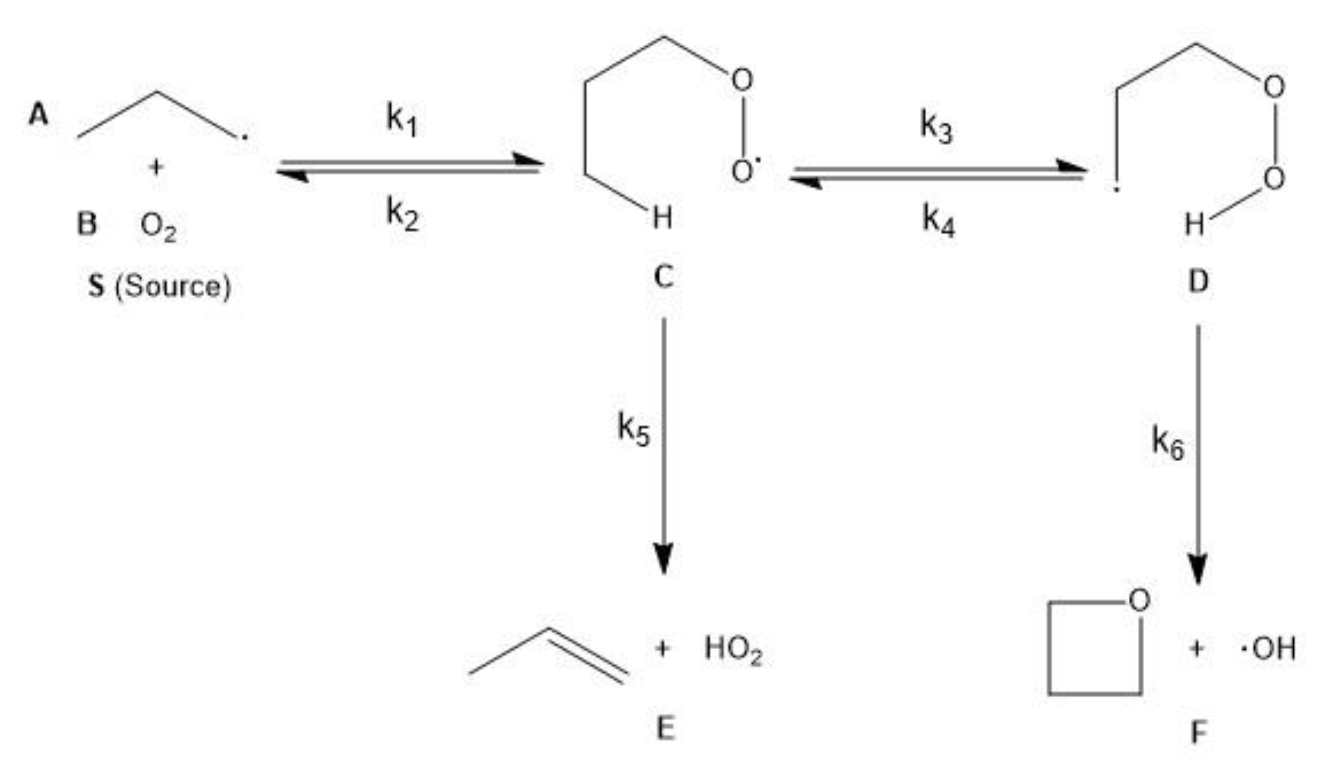}
\caption{The expanded scheme}
\centering
\end{figure}
The formations of \textbf{E} and \textbf{F} are treated as irreversible. This is because a stable alkene and a metastable hydroperoxyl are formed when forming products E, while the formation of products F cleaves the unstable peroxyl bond. The task, then, is to construct a matrix to account for all the reactions. Since these reactions can perturb the populations of the energised levels of the two isomers, they can also affect the rate constants deduced for isomerization, so it is important to include all the possibilities.

Combination reactions between radicals are often nearly barrierless, meaning that the rate constants are approximately temperature-independent and that the reverse dissociation reaction does not have an obvious transition state. However, it is possible to deduce the microcanonical dissociation rate constant from the high pressure-limiting (canonical) association rate constant.
$k_1$ can be modelled using an extended Arrhenius expression.
\begin{equation} \label{eq:a}
k_{1}=A_{a} \beta^{-v} e^{-\beta E_{a}}
\end{equation}
If $E_{a} \approx 0$, then the exponential term approximately equals 1, so that
\begin{equation} \label{eq:b}
k_{1}=A_{a} \beta^{-v}
\end{equation}
This form will be useful in determining $A_{a}$ and $v$.

This reaction can be modelled using a simple model proposed by Gorin et.al \cite{Gorin}. The model is based on a thermal average of the centrifugal barrier with an attractive London potential. It can be shown that
\begin{equation} \label{eq:c}
A_{a}=\kappa 2^{\frac{3}{2}} 3^{\frac{1}{2}} \Gamma\left(\frac{2}{3}\right) \pi^{\frac{1}{2}}\left[\alpha_{A} \alpha_{B}\left(\frac{I_{A} I_{B}}{I_{A}+I_{B}}\right)\right]^{\frac{1}{3}}
\end{equation}

depending on the polarizabilities, the ionisation potentials, and the transmission coefficient, and
\begin{equation} \label{eq:d}
v=\frac{1}{6}
\end{equation}

$k_2$ can be determined using a model proposed by Davies et.al \cite{Davies}. Since the equilibrium constant is the ratio of the high-pressure limiting rate constants and can also be related to the partition functions,
\begin{equation} \label{eq:e}
k_{2}^{\infty}(T)=\frac{\left(\frac{q_{A}}{V}\right)\left(\frac{q_{B}}{V}\right)}{\frac{q_{C}}{V}} e^{-\frac{\Delta E_{0}}{R T}} k_{1}^{\infty}(T)
\end{equation}

As
\begin{equation} \label{eq:f}
k_{2}^{\infty}(T)=\frac{1}{q_{C, \mathrm{vr}}(T)} \int_{0}^{\infty} \rho_{\mathrm{A}, \mathrm{vr}}(E) k_{2}(E) e^{-\beta E} d E
\end{equation}

where vr stands for vibration and rotation, it can be shown, via an Inverse Laplace Transformation, that if $E<\Delta E_{0}+E_{a}$,
\begin{equation} \label{eq:g}
k_{2}(E)=0
\end{equation}
if $E \geq \Delta E_{0}+E_{a}$
\begin{equation} \label{eq:h}
k_{2}(E)=\frac{A_{a}^{\infty} C}{\Gamma\left(v+\frac{3}{2}\right) \rho_{C, \mathrm{vr}}(E)} \int_{0}^{E-\left(\Delta E_{0}+E_{a}\right)} \rho_{\mathrm{P}, \mathrm{vr}}\left(E_{P}\right) E_v d E_{P}
\end{equation}
in which
\begin{equation} \label{eq:ha}
E_v = \left(E-\left(\Delta E_{0}+E_{a}+E_{P}\right)\right)^{v+\frac{1}{2}}
\end{equation}

Both $k_5$ and $k_6$ can be calculated via RRKM routines with internal rotor corrections, as is outlined before. The new set of coupled Master Equations for this extended scheme is then
\begin{equation} \label{eq:i}
\frac{d \boldsymbol{p}^{C}}{d t}=\left(\omega^{C}\left(\boldsymbol{P}^{C}-\boldsymbol{I}^{C}\right)-\boldsymbol{K}^{\mathrm{CS}}-\boldsymbol{K}^{\mathrm{CD}}-\boldsymbol{K}^{\mathrm{CE}}\right) \boldsymbol{p}^{C}+\boldsymbol{K}^{\mathrm{DC}} \boldsymbol{p}^{D}+k^{\mathrm{SC}} p^{S} \boldsymbol{f}
\end{equation}
\begin{equation} \label{eq:j}
\frac{d \boldsymbol{p}^{D}}{d t}=\left(\omega^{D}\left(\boldsymbol{P}^{D}-\boldsymbol{I}^{D}\right)-\boldsymbol{K}^{\mathrm{DC}}-\boldsymbol{K}^{\mathrm{DF}}\right)\boldsymbol{p}^{D}+\boldsymbol{K}^{\mathrm{CD}} \boldsymbol{p}^{C}
\end{equation}
\begin{equation} \label{eq:jb}
\frac{d p^{S}}{d t}=\left|\boldsymbol{k}^{C S} \boldsymbol{p}^{C}\right|-k^{\mathrm{SA}} p^{S}
\end{equation}

where $\mathrm{S}$ stands for the source that produces the peroxyl radical, and $\boldsymbol{f}$ is the normalised energy distribution for $\mathrm{RO}_{2}$ at the moment of its formation, i.e., the normalised energy distribution of the source rate.

Suppose the rate of formation of $\mathrm{RO}_{2}$ at energy $E$ from the reaction is
\begin{equation} \label{eq:k}
k_{1}^{\infty}[R]\left[O_{2}\right] f(E)
\end{equation}

If the reaction reaches equilibrium, then this rate would be
\begin{equation} \label{eq:l}
k_{1}^{\infty}[R]_{e q}\left[O_{2}\right]_{e q} f(E)
\end{equation}

and the energy distribution is the same, because it just depends on A and B colliding in a Boltzmann distribution.

At equilibrium, detailed balance is obeyed, and the product is also in a Boltzmann distribution, i.e.
\begin{equation} \label{eq:m}
k_{1}^{\infty}[R]_{e q}\left[O_{2}\right]_{e q} f(E)=k_{2}(E) b_{\mathrm{RO}_{2}}(E)\left[\mathrm{RO}_{2}\right]_{e q}
\end{equation}

in which $b_{\mathrm{RO}_{2}}(E)$ is a normalised Boltzmann distribution in $\mathrm{RO}_{2}$ :
\begin{equation} \label{eq:n}
b_{\mathrm{RO}_{2}}(E)=\frac{\rho_{\mathrm{vr}, \mathrm{RO}_{2}}(E) e^{-\beta E}}{q_{\mathrm{vr}, \mathrm{RO}_{2}}}
\end{equation}

By using the equilibrium constant, $f(E)$ can be isolated:
\begin{equation} \label{eq:o}
f(E)=\frac{K_{c}}{k_{1}^{\infty}} k_{2}(E) b_{\mathrm{RO}_{2}}(E)=\frac{k_{2}(E)}{k_{2}^{\infty}} b_{\mathrm{RO}_{2}}(E)
\end{equation}

where $k_{2}^{\infty}$ is the high-pressure limiting dissociation rate constant.

An extended matrix can therefore be constructed from these coupled Master Equations by considering the source. This is done by adding a new row and a new column at the left and top of the matrix. A table-like demonstration that accounts for the new structure $\mathbf{M}^{\prime}$ and vector $\mathbf{p}^{\prime}$ is shown in Figure 7.
\begin{figure}[H]
\includegraphics[width=\linewidth]{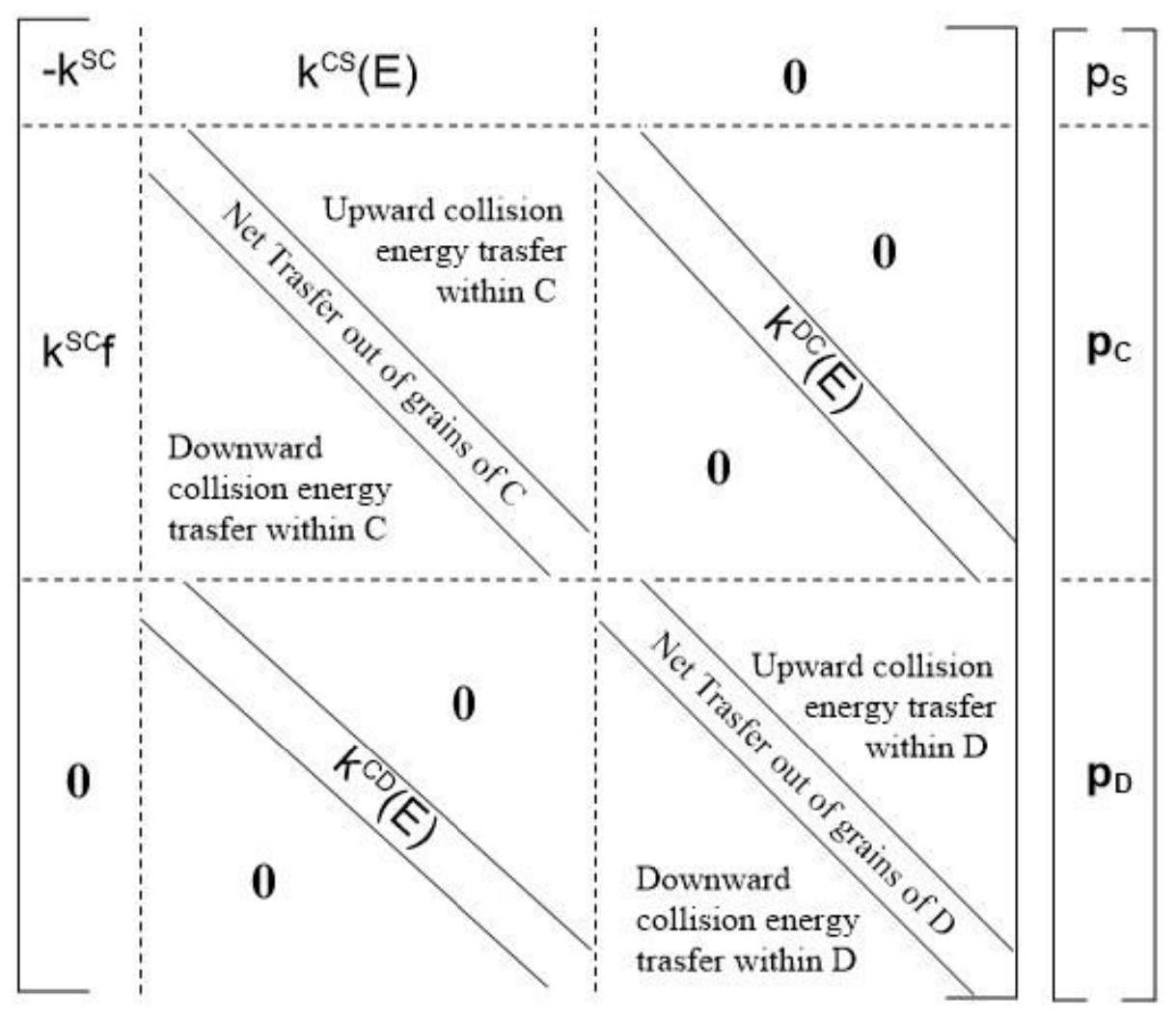}
\caption{The extended matrix}
\centering
\end{figure}

\subsection{Relevant Experimental Results and Explanation}

Similarly, changes are going to be observed in the overall relaxation behaviour after including source and depletion. Figure 8 below shows the change in the relaxation rate constant at 600K before and after including the source and drain term (reaction to form smaller molecules).
\begin{figure}[h]
\includegraphics[width=\linewidth]{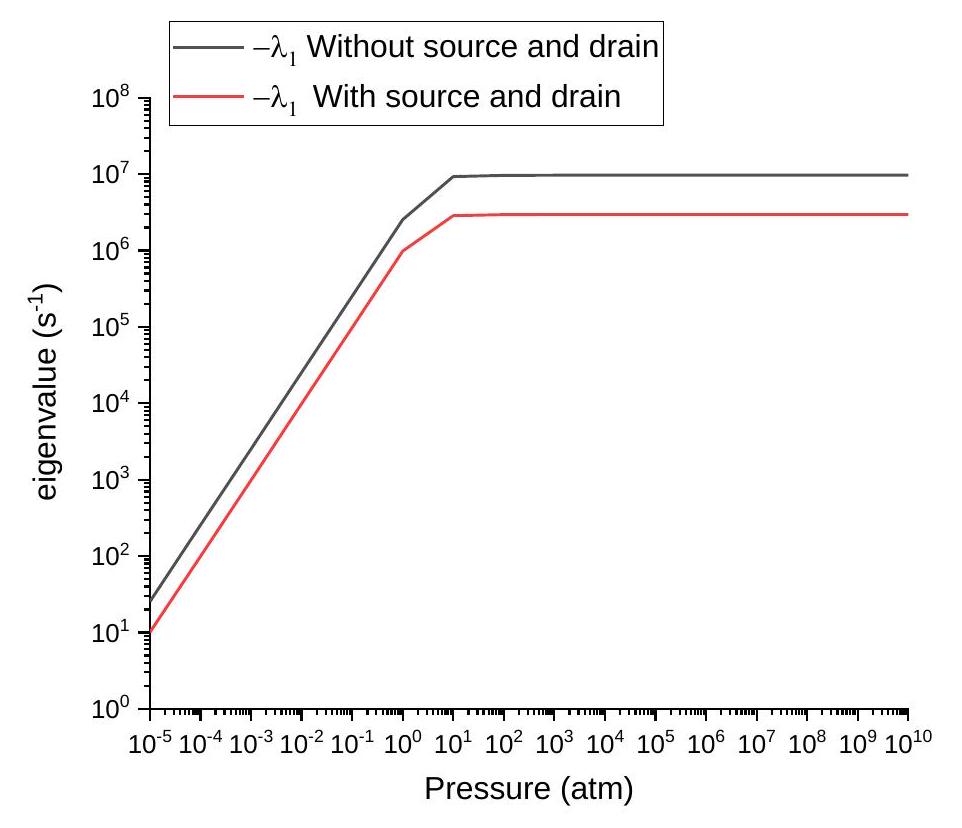}
\caption{The change of the relaxation rate constant with respect to source and drain at 600K}
\centering
\end{figure}

As is shown in the graph, the value of $-\lambda_{1}$ is reduced at all pressures when the source term and the drain term are included. This result is rather intuitive, as constantly providing high-energy reactant $\mathbf{C}$ slows the system down from reaching equilibrium.

One can also investigate how rate constants depend on temperature by changing the value of temperature in the program Figure 9 displays the expected outcome of the relationship between the pressure and the values of rate constants at various temperatures.
\begin{figure}[H]
\includegraphics[width=\linewidth]{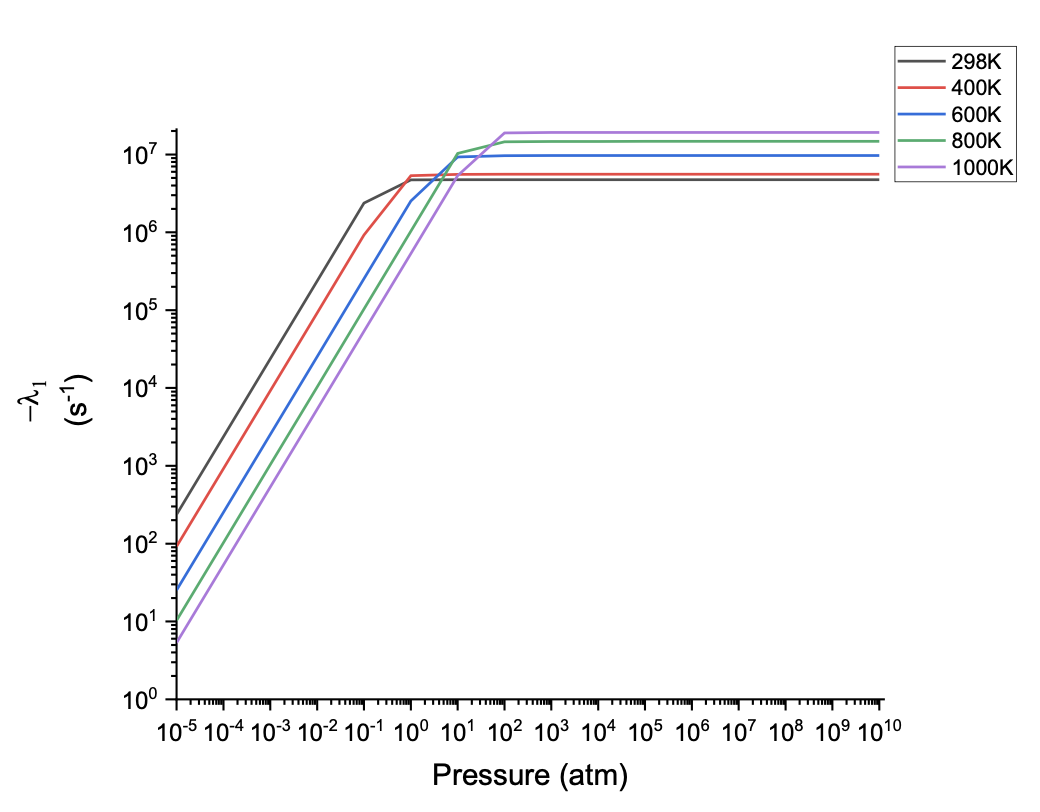}
\caption{The change of the relaxation rate constant with respect to temperature at 600K}
\centering
\end{figure}

When the temperature is low, the rate constants will be faster at low pressures, according to the graph. However, the reverse trend is seen at high pressure. To explain this, it should be noted that at low pressure values, the deficiency of collisions means that the Boltzmann distribution does not hold since there is not much energy exchange. The result of this is that the collision frequency will be the dominant term in determining the rate constant. Its value is then lower at high temperatures due to the fact that collision frequency is inversely proportional to temperature at constant pressure.

At high pressure values, however, the sufficient number of collisions means that the energy distribution can be approximated as Boltzmann. We can interpret the system Using the canonical TST again, which implies that the rate constants at the same pressure increase with temperature.

The tunnelling effect also has a profound impact on rate constants. Because this is more evident at low pressures and low temperatures, the ratio of the rate constants between 298K and 1000K at $10^{-5}$ Pa should be larger when the quantum effect is included. This is clearly demonstrated by the programme, as the ratio experiences a twelve-fold increase from 4 to 47 after ‘turning on’ the tunnelling. It is worth mentioning that the rate constants change little at 1000K regardless of quantum effects, which is consistent with the fact that tunnelling is negligible at high temperatures.

%%%%%%%%%%%%%%%%%%%%%%%%%%%%%%%%%%%%%%%%%%%%%%%%%%%%%%%%%%%%%%
\section{Conclusion}
%%%%%%%%%%%%%%%%%%%%%%%%%%%%%%%%%%%%%%%%%%%%%%%%%%%%%%%%%%%%%%
\

In this tutorial, a step-by-step guide has been written to calculate the rate constant of a free radical isomerization reaction that goes through a cyclic transition state, including the basics of RRKM theory, the relationship between the density of states and the partition function, the reversible ME, as well as more advanced ones such as considering tunnelling, hindered rotors, and source and depletion. To conclude, this tutorial provides a meaningful guide on how to construct a Master Equation model to study a chemical reaction from scratch. It is expected that one can acquire the key basics of this methodology from the sections and further expand it to study more complex systems and reaction schemes. 

%%%%%%%%%%%%%%%%%%%%%%%%%%%%%%%%%%%%%%%%%%%%%%%%%%%%%%%%%%%%%%
\section*{Acknowledgements}
%%%%%%%%%%%%%%%%%%%%%%%%%%%%%%%%%%%%%%%%%%%%%%%%%%%%%%%%%%%%%%
The author acknowledges the University of Chicago for providing the teach assistant fellowship for this research. The author also thanks Nicholas Green for his careful guidance through this project and Martin Galpin for providing meaningful discussions.

%%%%%%%%%%%%%%%%%%%%%%%%%%%%%%%%%%%%%%%%%%%%%%%%%%%%%%%%%%%%%%
%\section*{References}

%%%%%%%%%%%%%%%%%%%%%%%%%%%%%%%%%%%%%%%%%%%%%%%%%%%%%%%%%%%%%%

\end{document}